\input{aipcheck}

\documentclass[
    ,final            
  ]
  {aipproc}

\layoutstyle{6x9}

\begin{document}

\title{GRB Host Studies (GHostS)}


\author{S.\ Savaglio}{
  address={Johns Hopkins
University, Baltimore, USA}
}

\author{K.\ Glazebrook}{
  address={Johns Hopkins
University, Baltimore, USA}
}

\author{D. Le Borgne}{
  address={CEA/Saclay, France}
}

\begin{abstract}
The GRB Host Studies (GHostS) is a public archive collecting observed
quantities of GRB host galaxies.  At this time (January 2006) it
contains information on 32 GRB hosts, i.e.\ about half of the total
number of GRBs with known redshift. Here we present some preliminary
statistical analysis of the sample, e.g.\ the total stellar mass,
metallicity and star formation rate for the hosts. We found that these
are generally low-mass objects, with 79\% having $M_\ast<10^{10}$
M$_\odot$. The total stellar mass and the metallicity for a subsample
of 7 hosts at $0.4<z<1$ are consistent with the mass-metallicity
relation recently found for normal star-forming galaxies in the same
redshift interval. At least 56\% of the total sample are bursty
galaxies: their growth time-scale (the time required to form the
observed stellar mass assuming that the observed SFR is constant over
the entire life of the galaxy) is shorter than 400 Myr.

\end{abstract}

\maketitle


\section{Introduction}

Although the typical nature of GRB hosts is still heavily under
discussion, it is clear that they differ from normal high-$z$
galaxies, as they are generally low-luminosity and young objects
\cite{Emeric,Christ}.  We still cannot tell whether they form a galaxy
population by themselves, or they are just much easier to detect than
normal low-luminosity galaxies because they are associated with
transient, but very luminous events.

To help investigate this issue, we have initiated a database dedicated
to GRB host galaxies, called GRB Host Studies\footnote{GHostS can be
accessed at the URL http://www.pha.jhu.edu/$\sim$savaglio/ghosts}
(GHostS). Thanks to the advent of the Swift mission, the amount of
results related to GRB hosts is in rapid ascent. The goal of GHostS is
to gather, classify and synthesize GRB host information, and derive
meangful parameters for a new statistically significant sample. At the
present, it is the largest public archive of its kind.  For each host,
the optical-NIR photometry is provided, together with emission line
fluxes, originating in the star-forming regions. So far, GHostS uses
results coming from more than 70 different publications.

In this first work, we focus on the determination of the total stellar
mass of the host galaxies, a parameter that has been hardly
investigated in the past, for a number of good reasons \cite{Chary}. We
relate stellar masses to SFRs and metallicities.

\section{The GRB host sample}

\begin{figure}  
\includegraphics[height=.4\textheight]{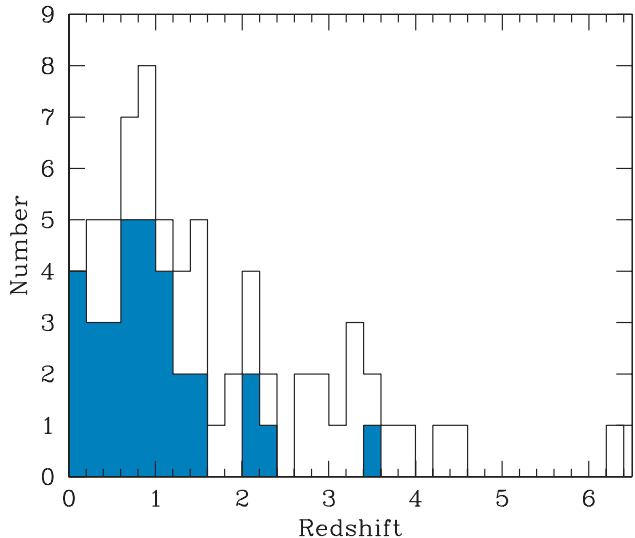} 
\caption{Redshift histogram of all GRBs with known redshift
(67 in total - {\it open histogram}) and of the 32 GRBs with detected
host, currently included in the GHostS archive ({\it filled
histogram}). The median redshift of the two samples is $z\simeq1.12$
and $z\simeq0.84$, respectively.}
\label{f1}
\end{figure}

The median redshift of our 32 GRB hosts is $z\simeq0.84$
(Figure~\ref{f1}), i.e.\ lower than the present median redshift for
all GRBs with measured redshift ($z\simeq1.12$). A couple of them are
associated with a short-duration GRB, the remaining are long-duration
GRBs. The objects are selected according to the requirement that
optical-NIR photometry (necessary to estimate the total stellar mass)
is available.  Optical-NIR photometry is hard to measure for $z>2$
galaxies in general. Our sample contains 4 objects with $z>2$.

Among the 32 hosts, fluxes of [OII], [OIII] and H$\beta$ emission
lines are available for 19, 10 and 9 of them, respectively. These are
used to derive metallicities and SFRs.

\begin{figure}  
\includegraphics[height=.4\textheight]{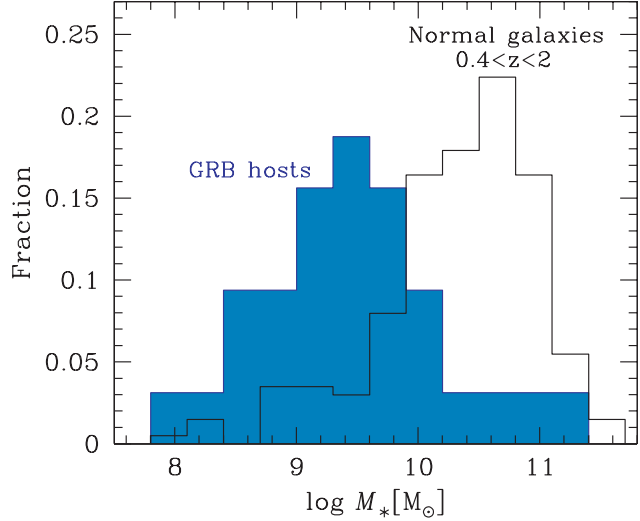}
\caption{Fraction of GRB hosts per stellar-mass bins ({\it filled 
histogram}) and the comparison with 201 normal $0.4<z<2$ galaxies
({\it open histogram}) from the $K<20.6$ Gemini Deep Deep Survey (see
also \cite{Karl}).  In the GRB host sample, 79\% have
stellar masses below $M_\ast=10^{10.0}$ M$_\odot$. The GDDS is
complete for stellar masses above $M_\ast=10^{10.8}$ M$_\odot$ and
$M_\ast=10^{10.1}$ M$_\odot$, for all galaxies and star-forming
galaxies, respectively.}
\label{f2}
\end{figure}

\subsection{Total stellar masses}

The stellar mass of the hosts is derived using a procedure which will
be described in detail in a future paper (Le Borgne, Glazebrook \&
Savaglio, in preparation). Briefly, our technique uses SED fitting to
the multi-band optical-NIR photometry.  The observed NIR light, which
in high-$z$ galaxies samples rest-frame light above the 4000\AA\
break, is closely related to the galaxy's total stellar mass and
provides good stellar mass estimates up to $z=2-3$ \cite{Karl}. It is
also insensitive to dust because most stars in a galaxy are not in
birth clouds, and because the redder bands are less affected by
extinction. The stellar mass derived this way is a much more
meaningful physical variable than the luminosity, because it
represents the integral of the past star-formation and merger history,
and, in contrast for instance to UV light, can only increase with
time.

In Figure~\ref{f2} we show the stellar mass histogram for the 32 GRB
hosts.  The median/average mass and 1$\sigma$ dispersion are
$M_\ast=10^{9.5}$ M$_\odot$ and 0.9 dex, respectively. This is
compared to the same histogram obtained for normal $0.4<z<2$ galaxies
from the Gemini Deep Deep Survey (GDDS; \cite{Bob}). The
GDDS is a deep optical-NIR ($K<20.6$) survey and is complete at
$0.4<z<2$ for all galaxies and for star-forming galaxies down to
stellar masses $M_\ast= 10^{10.8}$ M$_\odot$ and $M_\ast=10^{10.1}$
M$_\odot$, respectively. The comparison shows that GRB observations
are much more efficient in detecting low-mass galaxies at high
redshift than traditional high-$z$ surveys.

\subsection{Metallicities and SFRs}

\begin{figure}
  \includegraphics[height=.4\textheight]{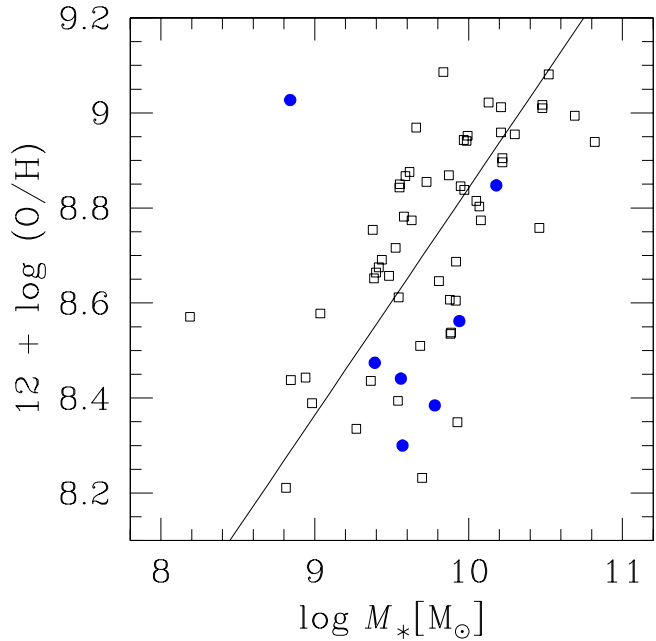} 
\caption{Total stellar mass and metallicity for $0.4<z<1$ 
GRB hosts ({\it filled circles}). Metallicities are derived assuming a
modest (10\%) Balmer stellar absorption correction and $A_V=1$ dust
extinction (Milky Way extinction law). The outlier at $\log
M_\ast=8.8$ is the GRB~991208 host galaxy at $z=0.706$
\cite{Alberto}. {\it Open squares}: results for $0.4<z<1$
galaxies from GDDS and CFRS \cite{Sandra}. The straight line
is the bisector fit for this sample. }
\label{f3}
\end{figure}

Metallicities are derived with the $R_{23}$ calibrator, which uses
[OII], [OIII] and H$\beta$ line fluxes \cite{Pagel}. We adopted the
formulation recently proposed by Kobulnicky \& Kewley (2004)
\cite{KK}. This set of emission lines allows the metallicity
measurement for the largest possible number of GRB hosts. The $12+\log
(\rm O/H)$ value is derived for 9 hosts, 7 of which are in the
redshift interval $0.4<z<1$ (Figure~\ref{f3}). Figure~\ref{f3} also
shows the same parameters derived for GDDS and Canada-France Redshift
Survey\footnote{Emission line fluxes for the CFRS galaxies are taken
from Lilly et al.\ (2003) \cite{Lilly}} (CFRS) galaxies, in the same redshift
interval \cite{Sandra}.

We estimate SFRs using the [OII] emission, the most common line
measured in GRB hosts. The H$\alpha$ emission flux provides a more
robust SFR estimate, however this is measured in 5 GRB hosts
only. Moustakas et al.\ (2006) \cite{JohnM} have shown that the
dust-corrected [OII]-to-H$\alpha$ flux ratio in local galaxies is on
average one (0.12 dex dispersion) over a large range of $B$
luminosities (from $10^7$ to $10^{11}$ L$_\odot$). We adopt this
relation to derive SFRs, after assuming a Milky Way extinction law
with $A_V=1$. The median SFR and the range spanned by the 19 GRB hosts
are SFR $=12$ M$_\odot$ yr$^{-1}$ and $1-100$ M$_\odot$ yr$^{-1}$,
respectively (we also apply a correction of a factor of 2 for
slit-aperture loss).

\begin{figure}
  \includegraphics[height=.4\textheight]{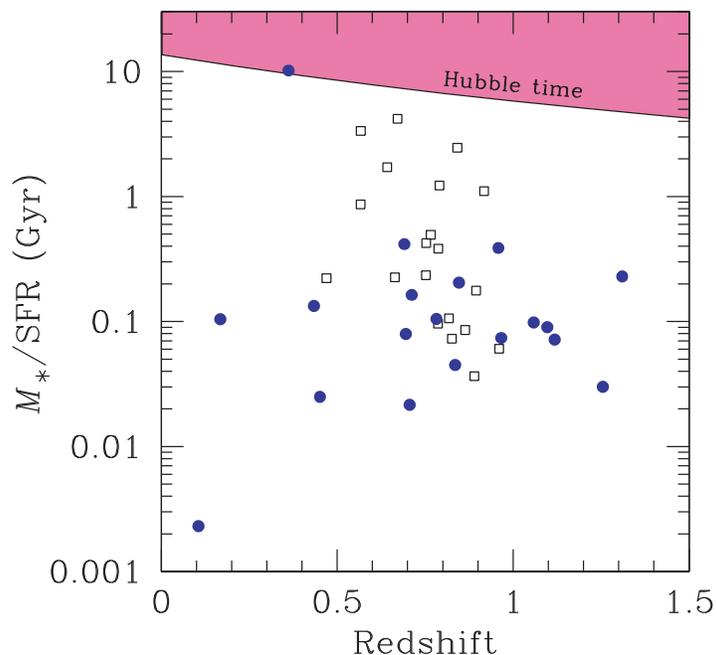} 
\caption{Growth time-scale (total stellar mass divided by observed SFR)
as a function of redshift for 19 GRB hosts ({\it filled dots}). These
are compared to the $0.4<z<1$ star-forming galaxies of GDDS and CFRS
of Figure~\ref{f3} ({\it open squares}). The curve marks the age of
the universe (Hubble time) as a function of redshift, and indicates
the transition from the quiescent star-formation mode to the bursty
star-formation mode.}
\label{f4}
\end{figure}

Another indicative parameter is the SFR per mass unit, or the inverse
of this, also called growth time-scale. This is defined as $\rho_\ast
\equiv M_\ast/{\rm SFR}$ \cite{Stephanie}, and is the time that a
galaxy needs to build the observed stellar mass, if the SFR is assumed
to be constant and is given by the observed value.  The result as a
function of redshift is shown in Figure~\ref{f4}, together with the
comparison with normal star-forming $0.4<z<1$ galaxies from GDDS and
CFRS \cite{Sandra}.
 
\section{Results}

We presented the first results of the GHostS project. We focused on
the stellar mass of 32 GRB hosts, SFRs for a subsample of 19, and
metallicities for a subsample of 9, and compared these to normal
high-$z$ galaxies. We found that the total stellar mass is
$<10^{10.0}$ M$_\odot$ for 79\% of the GRB hosts (Figure~\ref{f2}).
At this mass, most of the high-$z$ deep spectroscopic surveys are
highly incomplete. The median stellar mass of the sample is
$M_\ast=10^{9.5}$ M$_\odot$, i.e.\ comparable to the stellar mass
content of the Large Magellanic Cloud. Recently, Chary et al.\ (2002)
\cite{Chary} have derived stellar masses for 7 hosts and found a 
similar median value.

The median observed and dust-corrected (for $A_V=1$) SFR in 19 hosts
are 3 and 12 M$_\odot$ yr$^{-1}$, respectively. Given the generally
low stellar masses for these GRB hosts, we conclude that a large
fraction are bursty galaxies, with growth time-scales that are shorter
than 400 Myr, and on average 100 Myr (Figure~\ref{f4}). If we consider
the whole GRB population with measured redshift (67 in total), the
host is a bursty galaxy in at least 1/4 of the cases.

The median metallicity in 9 GRB hosts is 0.6 solar, with values
ranging from half to twice solar.  These values are not far from
expectations, given the stellar masses and redshifts of the galaxies.
Moreover, they behave as predicted by the mass-metallicity relation
observed at high redshifts in normal star-forming galaxies
(Figure~\ref{f3}).

In summary, we quantified some of the known statements regarding GRB
hosts, according to which a large fraction of them are low-mass
starbursts. Although low-mass starbursts at high redshifts are hard to
identify if no GRB event occurs, there is no evidence that GRB hosts
represent a different population of galaxies that existed in the young
universe.

\begin{theacknowledgments}
The authors thank the conference organizers for offering such an
interesting event and enjoyable time, and the Swift team for their large
contribution to our understanding of GRB phenomenology.
\end{theacknowledgments}


\begin{thebibliography}{12}
\bibitem{Emeric}
Le Floc'h, E., et al.\ 2003, A\&A, 400, 499 
\bibitem{Christ}
Christensen, L., Hjorth, J., \& Gorosabel, J.\ 2004, A\&A, 425, 913
\bibitem{Chary} 
Chary, R., Becklin, E. E., Armus, L.\ 2002, ApJ, 566, 229
\bibitem{Karl}
Glazebrook, K., et al.\ 2004, Nature, 430, 181 
\bibitem{Bob} 
Abraham, R.\ G.\ et al.\ 2004, AJ, 127, 2455
\bibitem{Pagel}
Pagel, B.~E.~J., Edmunds, M.~G., Blackwell, D.~E., Chun, M.~S., \& Smith, G.\ 1979, MNRAS, 189, 95
\bibitem{KK}
Kobulnicky, H.~A., \& Kewley L.~J.\ 2004, ApJ, 617, 240
\bibitem{Lilly}
Lilly, S. J., Carollo, C. M., \& Stockton, A. N. 2003, ApJ,  597, 730
\bibitem{Alberto}
Castro-Tirado, A.~J., et al.\ 2001, A\&A, 370, 398 
\bibitem{Sandra}  
Savaglio, S., et al.\ 2005, ApJ, 635, 260
\bibitem{JohnM}
Moustakas, J., Kennicutt, R.~C., Tremonti, C.~A.\ 2006, ApJ, in press (astro-ph/0511730)
\bibitem{Stephanie}
Juneau, S., et al.\ 2005, ApJL, 619, 135

\end{thebibliography}
\end{document}